\documentclass{article}
\usepackage{FPSpr1}
\input psfig.sty

\begin{document}

\def\lcdm{$\Lambda$CDM}

\def\ltsima{$\; \buildrel < \over \sim \;$}
\def\lsim{\lower.5ex\hbox{\ltsima}}
\def\gtsima{$\; \buildrel > \over \sim \;$}
\def\gsim{\lower.5ex\hbox{\gtsima}}

\def\kms{{\rm km}\, {\rm s}^{-1}}
\def\kmsMpc{{\rm km}\, {\rm s}^{-1}\,{\rm Mpc}^{-1}}
\def\hMpc{h^{-1}{\rm Mpc}}
\def\Msun{M_{\odot}{\ }}
\def\hMsun{h^{-1}M_{\odot}{\ }}
\def\Mvir{M_{\rm vir}}
\def\mvir{M_{\rm vir}}
\def\rvir{R_{\rm vir}}
\def\Cvir{c_{\rm vir}}
\def\cvir{c_{\rm vir}}
\def\Rs{r_{\rm s}}
\def\rs{r_{\rm s}}
\def\Rvir{R_{\rm vir}}
\def\rvir{R_{\rm vir}}
\def\vmax{V_{\rm max}}
\def\Vmax{V_{\rm max}}
\def\vmax{V_{\rm max}}
\def\nwm2sr{nW m$^{-2}$ sr$^{-1}$}

\title{PROBING GALAXY FORMATION WITH HIGH ENERGY GAMMA-RAYS}
\author{Joel R. Primack\\
   {\em Physics Department, University of California, Santa Cruz, CA
95064 USA}}

\maketitle
\baselineskip=11.6pt
\begin{abstract}
I discuss how measurements of the absorption of $\gamma$-rays from GeV
to TeV energies via pair production on the extragalactic background
light (EBL) can probe important issues in galaxy formation.  My group
uses semi-analytic models (SAMs) of galaxy formation, set within the
CDM hierarchical structure formation scenario, to obtain predictions
of the EBL from 0.1 to 1000$\mu$m. SAMs incorporate simplified
physical treatments of the key processes of galaxy formation ---
including gravitational collapse and merging of dark matter halos, gas
cooling and dissipation, star formation, supernova feedback and metal
production --- and have been shown to reproduce key observations at
low and high redshift.  We have improved our modelling of the spectral
energy distributions in the mid-to-far-IR arising from emission by
dust grains.  Assuming a flat \lcdm\ cosmology with $\Omega_m=0.3$ and
Hubble parameter $h=0.65$, we investigate the consequences of
variations in input assumptions such as the stellar initial mass
function (IMF) and the efficiency of converting cold gas into stars.
We also discuss recent attempts to determine the emitted spectrum of
high energy gamma rays from blazars such as Mrk 501 using the
synchrotron self-Compton model and the observed X-rays, and note that
our favorite SAM EBL plus the observed spectrum of Mrk 501 do {\it
not} imply unphysical upturns in the high energy emitted spectrum ---
thus undermining recent claims of a crisis with drastic possible
consequences such as breaking of Lorentz invariance.  We conclude that
observational studies of the absorption of $\gamma$-rays with energies
from $\sim$10 Gev to $\sim$10 TeV will help to determine the EBL, and
also help to explain its origin by constraining some of the most
uncertain features of galaxy formation theory, including the IMF, the
history of star formation, and the reprocessing of light by
dust.\footnote{This paper is an updated version of
\protect\cite{PGam2000}.}
\end{abstract}
\baselineskip=14pt

\section{Introduction}

The extragalactic background light (EBL) represents all the light that
has been emitted by galaxies over the entire history of the
universe. The EBL that we observe today is an admixture of light from
different epochs, its spectral energy distribution (SED) distorted by
the redshifting of photons as they travel to us from sources at
different distances. It is therefore a constraint on both the
intrinsic SEDs of the sources and their distribution in redshift.  At
present, there is more than a factor of two uncertainty in the
amplitude of the EBL in the UV, optical, and near-infrared
\cite{puget}. The EBL in the mid-IR is even more uncertain. The far-IR
background measured at $\gsim 100 \mu$m
\cite{puget96,guider97,hauser,fixsen} represents at least half of the
total energy in the EBL, yet the sources that produced it remain
uncertain.

High energy $\gamma$-ray astronomy promises to help resolve these
uncertainties by providing independent constraints on the EBL, in the
mid-IR with $E_\gamma$ in the $\sim 10$ TeV energy range, and in the
0.1-3 $\mu$m range with $E_\gamma \sim100$ GeV via the new
low-threshold instruments that will soon be available.  High energy
$\gamma$-rays from sources at cosmological distances are absorbed via
electron-positron pair production on the diffuse background of photons
that comprises the EBL. Thus, $\gamma$-ray observations of objects
with known redshift and intrinsic spectral shape will constrain the
EBL in these crucial wavelength regimes by measuring the optical depth
of the Universe to photons of various energies.  This in turn will
help to constrain some of the most fundamental uncertainties in
physical models of galaxy formation.

In order to illustrate this, in this paper we use a ``forward
evolution'' approach, which attempts to model the essential features
of galaxy formation using simple recipes. These semi-analytic models
are set within the modern Cold Dark Matter (CDM) paradigm of
hierarchical structure formation, and trace the gravitational collapse
and merging of dark matter halos, the cooling and shock heating of
gas, star formation, supernovae feedback, metal production, the
evolution of stellar populations and the absorption and re-emission of
starlight by dust. This machinery has been used extensively to predict
optical properties of low-redshift galaxies, with good results (e.g.,
\cite{kwg,cole94}; reviewed and extended in \cite{sp,spf}, hereafter
SP and SPF).  A semi-analytic approach was also used by Devriendt and
Guiderdoni \cite{dev2} to make predictions of counts and backgrounds
in the mid-to-far-IR, with more detailed modelling of dust extinction
and emission, but less detailed modelling of merging and star
formation.  We have now combined the strengths of these two approaches, 
by integrating the stellar SEDs and dust modelling of \cite{dev1,dev2}
into the galaxy formation SAM code of the Santa Cruz group.

Some parts of the ``standard paradigm'' of galaxy formation
represented by our SAMs are relatively solid. For example, once a
cosmological model and power spectrum are specified, it is
straightforward to compute the gravitational collapse of dark matter
into bound halos using $N$-body techniques, and analytic formalisms
such as those used in our modelling \cite{sk} have been checked
against these results \cite{slkd}. Within the range of values for the
cosmological parameters allowed by existing observational constraints
(i.e., $\Omega_{\rm matter} \simeq 0.3-0.5$, $\Omega_{\rm
matter}+\Omega_\Lambda \simeq 1$, $H_0 \simeq 60-80$ km/s/Mpc; see
e.g.~\cite{primack2000} for a summary), these results do not change
significantly.  Similarly, modelling of gas cooling appears to be
fairly robust and agrees well with hydrodynamic simulations
\cite{pearce}. However, other aspects, notably the efficiency of
conversion of cold gas into stars, the effect of subsequent feedback
due to supernovae winds or ionizing photons, the stellar initial mass
function (IMF), and the effects of dust, remain highly uncertain, and
some predictions are quite sensitive to their details.

For example, SPF showed that the star formation history of the
Universe and the number density of high redshift $z \gsim 2$
``Lyman-break'' galaxies (LBGs; e.g.~\cite{steidel:99}) may be quite
different depending on whether star formation is primarily regulated
by internal properties, such as gas surface density in a quiescent
disk, or triggered by an external event such as an interaction.
Because the largest samples of LBGs are primarily identified in the
rest UV, model predictions are also quite sensitive to the
high-stellar-mass slope of the IMF, and to dust extinction. At the
other end of the spectrum is the sub-mm population detected by SCUBA,
believed to be predominantly high redshift ($z \gsim 2$) 
luminous and ultraluminous infrared galaxies (LIRGs and ULIRGs)
powered by star formation rates of hundreds to thousands of solar
masses per year (e.g., \cite{sanders}). Theoretical predictions of the
numbers and nature of these objects are highly sensitive to the same
issues (the dominant mode of star formation, dust, the IMF), but
provide a crucial counter-balance to the optical observations. 
However, the current mismatch between the sensitivity and spatial
resolution of optical and sub-mm instrumentation has made it difficult
to establish the connection between the two populations
observationally.

The Milky Way, like most nearby galaxies, emits the majority of its
light in optical and near-IR wavelengths; only about 30\% of the
bolometric luminosity locally is released in the far-infrared
\cite{sn:91}.  This was generally believed to be typical of most of
the starlight at all redshifts until the discovery of the far-IR part
of the EBL by the DIRBE and FIRAS instruments on the COBE satellite,
at a level ten times higher than the no-evolution predictions based on
the local luminosity function of IRAS galaxies, and representing twice
as much energy as the optical background obtained from counts of
resolved galaxies \cite{madaupoz}. This result suggests that either
the dust extinction properties of ``normal'' galaxies change
dramatically with redshift, or a population of heavily extinguished
galaxies (perhaps analogous to local LIRGs and ULIRGs) is much more
common at high redshift than locally, or both.  Some of these galaxies
may have already been observed, at 15 $\mu$m by ISO \cite{elbaz99},
and at 850 $\mu$m by SCUBA \cite{blain}.

Guiderdoni et al.~\cite{ghbm,dev2} showed that their simplified
semi-analytic model could reproduce the multi-wavelength data only if
they introduced a population of heavily extinguished galaxies with
high star formation rates, and with strong evolution of number density
with redshift.  This population was introduced ad-hoc by
\cite{ghbm,dev2}, but as discussed by these authors, by \cite{silkdev}
(based on \cite{bss}), and also by SPF, the increasing importance of
starbursts at high redshift, due to the increasing merger rate and
higher gas fractions, is a natural mechanism to produce this
population. The models of SPF contain a detailed treatment of mergers
and the ensuing collisional starbursts, which has been calibrated
against the merger rate in cosmological $N$-body simulations
\cite{kolatt} and the starburst efficiency in hydrodynamical
simulations \cite{mh,somerville}.  Moreover, they produced good
agreement with observations of LBGs (e.g.~\cite{papovich}) and damped
Lyman-$\alpha$ systems (SPF and \cite{maller}) as well as low redshift
galaxies (SP). Therefore, it will be extremely interesting to see if
these same models, when combined with the more sophisticated treatment
of dust extinction and emission developed by Devriendt, Guiderdoni,
and collaborators, will be able to simultaneously reproduce
observations over the broad range of wavelengths and redshifts
discussed above.

In the next section we briefly describe the ingredients of our models,
and then present the results of the predicted EBL.  Section 4 presents
the implications for $\gamma$-ray attenuation, and \S5 briefly
discusses some alternative treatments and our own conclusions.  The
work summarized here is a brief, preliminary sample of the results
which will soon be presented in a series of papers, now in
preparation, on the EBL and its breakdown into various kinds of
sources and on the implications for $\gamma$-ray astronomy.

\section{Semi-analytic modelling}

In this section we briefly describe the ingredients of our
models. Readers can refer to SP and SPF for more details, and to
\cite{veritas} for a brief introduction.

Using the method described in \cite{sk}, we create Monte-Carlo
realizations of the masses of progenitor halos and the redshifts at
which they merge to form a larger halo. These ``merger trees'' reflect
the collapse and merger of dark matter halos within a specific
cosmology (each branching in the tree represents a halo merging event
--- for examples, see e.g.~\cite{wechsler01b}).  We truncate the trees
at halos with a minimum circular velocity of 40 km/s, below which we
assume that the gas is prevented from collapsing and cooling by
photoionization.  Each halo at the top level of the hierarchy is
assumed to be filled with hot gas, which cools radiatively and
collapses to form a gaseous disk. The cooling rate is calculated from
the density, metallicity, and temperature of the gas. Cold gas is
turned into stars using several simple recipes, depending on the mass
of cold gas present and the dynamical time of the disk. Supernovae
inject energy into the cold gas and may expell it from the disk and/or
halo if this energy is larger than the escape velocity of the
system. Chemical evolution is traced assuming a constant yield of
metals per unit mass of new stars formed. Metals are initially
deposited into the cold gas, and may later be redistributed by
supernovae feedback, and mixed with the hot gas or the diffuse
(extra-halo) inter-galactic medium.

When halos merge, the galaxies contained in each progenitor halo
retain their seperate identities either until they spiral to the
center of the halo due to dynamical friction and merge with the
central galaxy, or until they experience a binding merger with another
satellite galaxy orbiting within the same halo.  We take into account
subhalo truncation due to tidal effects in the larger halo.  All newly
cooled gas is assumed to initally collapse to form a disk, and major
(nearly equal mass) mergers result in the formation of a spheroid. New
gas accretion and star formation may later form a new disk, resulting
in a variety of bulge-to-disk ratios at late times.

For an assumed IMF, the stellar SED of each galaxy is then obtained
using stellar population models. Here we use the multi-metallicity
stellar SEDs of \cite{dev1} for the Salpeter and Kennicutt IMF cases,
and the solar metallicity GISSEL models \cite{bc} for the Scalo IMF.
(We have found that using evolving metallicity rather than solar
metallicity SEDs has a relatively small impact on the resulting EBL.)
Dust extinction is modelled using an approach similar to that of
\cite{dev2}. The optical depth of the disk is assumed to be
proportional to the column density of metals.
We then use a simple slab geometry where stars and gas are homogenously
mixed, and assign a random inclination to each galaxy to compute the
absorption. We use a metallicity dependent extinction curve, following
\cite{ghbm,dev2}.

\begin{figure} 
\noindent \begin{minipage}[t]{2.25in} \centering 
 \psfig{file=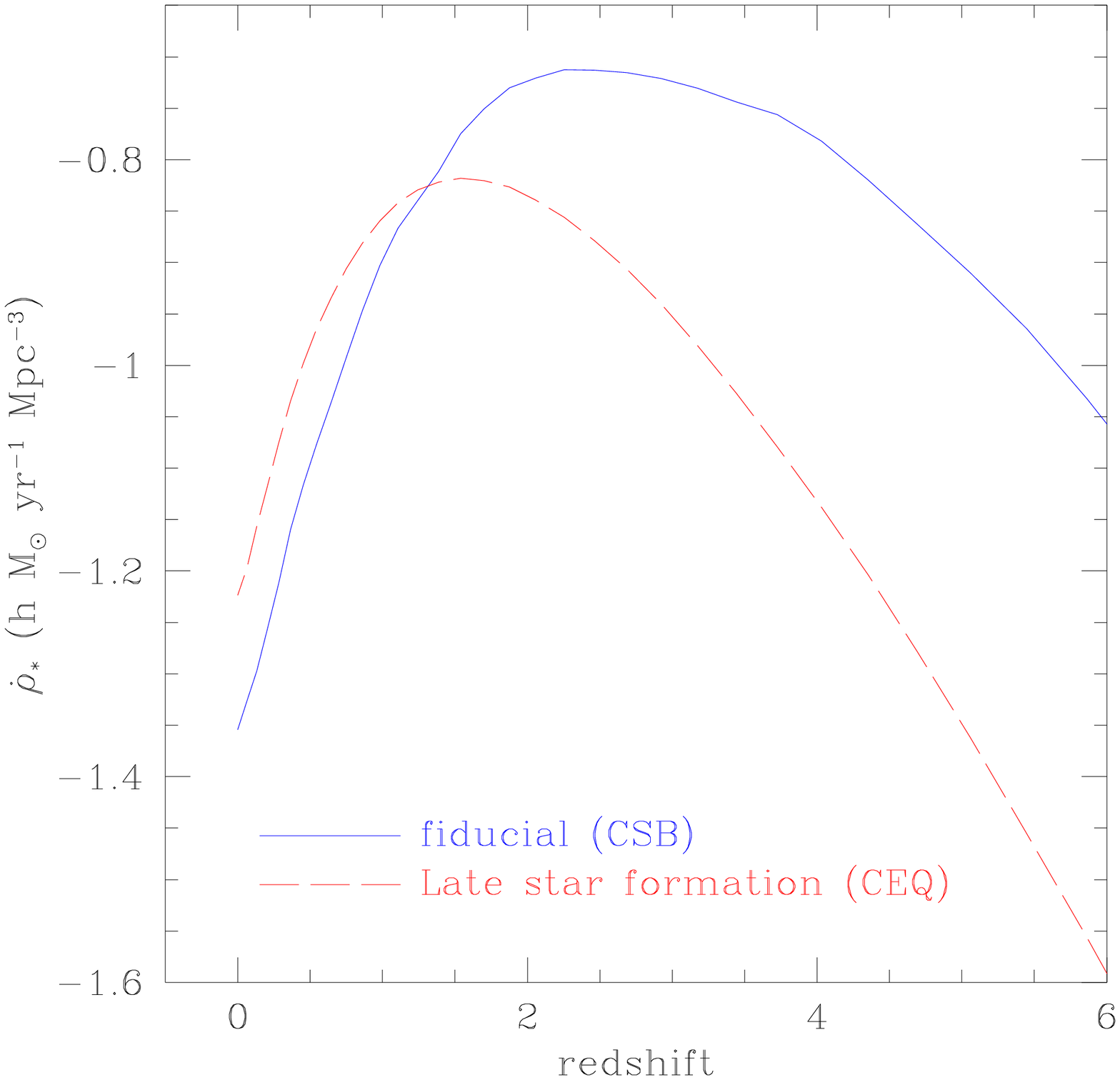,width=2.25in} \end{minipage} \hfill
\begin{minipage}[t]{2.25in} \centering 
 \psfig{file=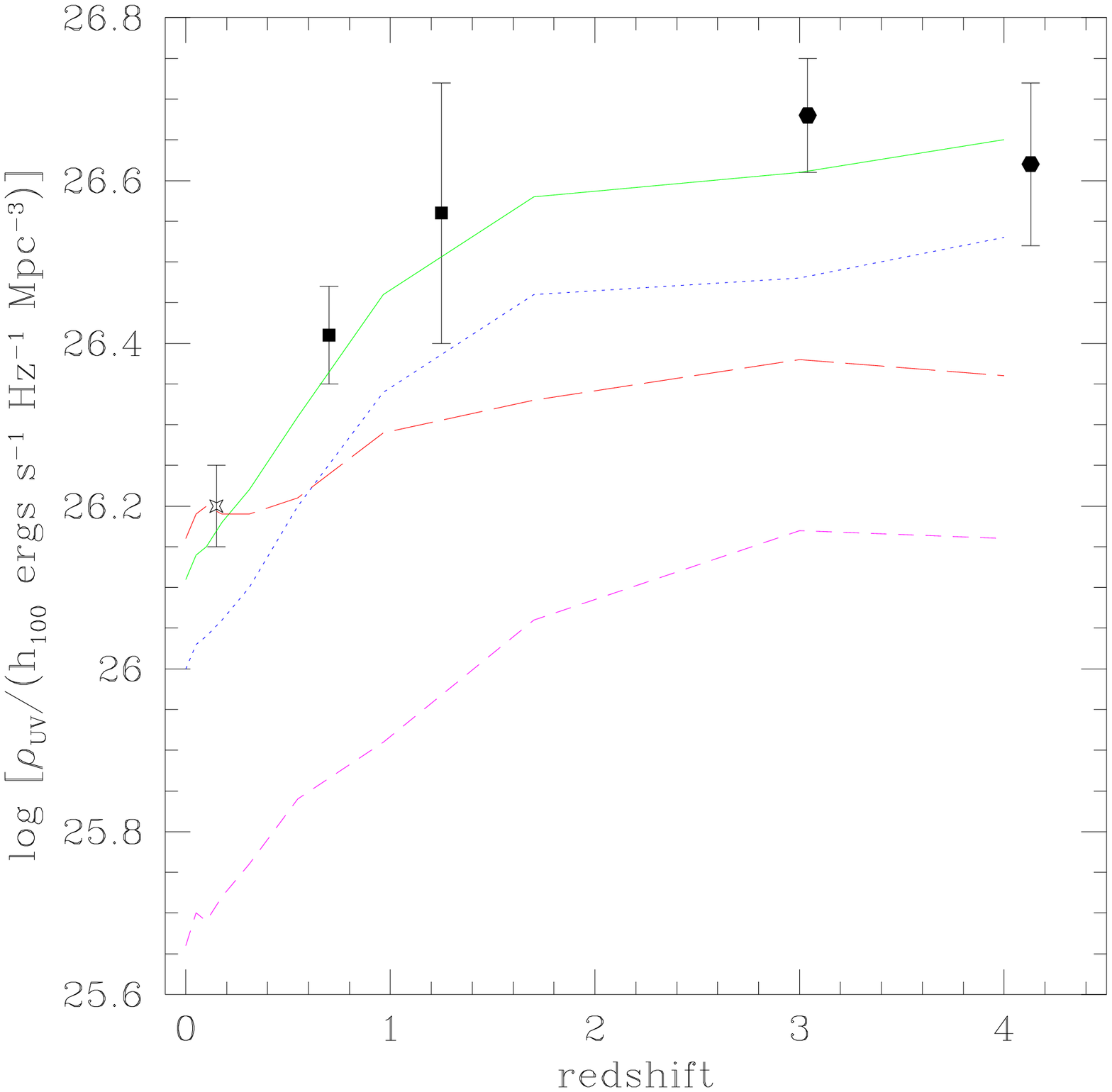,width=2.25in} \end{minipage} 
\caption{(a) The star formation rate density predicted by our models,
for two different recipes of star formation. Both models produce about
the same total mass density of stars by $z=0$ (i.e., the areas under
the curves are equal when they are plotted linearly vs. time), but the
collisional starburst model (CSB) peaks at higher redshift.  (b)
Comoving luminosity density at $2000 \AA$ as a function of redshift.
Data points represent the observed global luminosity density at rest
$\sim 2000 \AA$, obtained by integrating the observational best-fit
Schechter luminosity functions over all luminosities ($\rho_L = \phi_*
L_* \Gamma(2-\alpha)$), including corrections for dust extinction. The
$z=0.15$ point is from \protect\cite{sullivan}, the $z\sim 0.4$ and
1.2 points are from \protect\cite{cowie}, and the $z\sim3$ and
$z\sim4$ points are from \protect\cite{steidel:99}. The curves for our
four models are labeled as in Figure~\ref{fig:ebl}.  The model curves
have been corrected for dust extinction using the approach described
in the text.}
\label{UVlumdens}
\end{figure}

\begin{figure}[b!] 
\centerline{\psfig{file=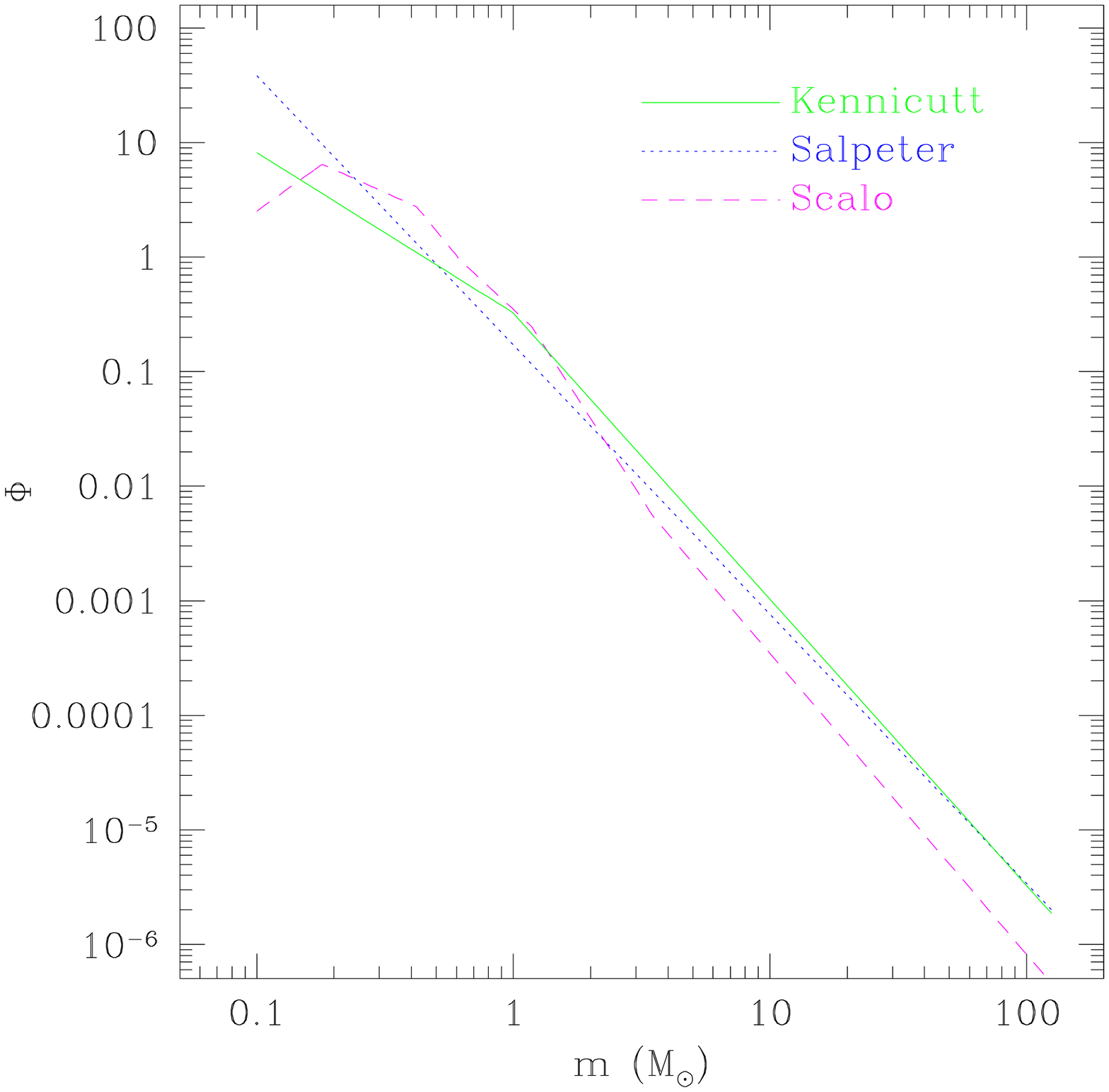,width=3in}}
\vspace{10pt}
\caption{The three stellar Initial Mass Functions (IMFs) used here:
Kennicutt \protect\cite{kenIMF}, Salpeter \protect\cite{salpeterIMF},
and Scalo \protect\cite{scaloIMF}.
}
\label{fig:imf}
\end{figure}

All absorbed light is re-radiated at longer wavelength.  The galactic
dust emission spectrum is represented by a combination of three
components: 1) hot dust (as in $H_{\rm II}$ regions), 2) warm dust (as
in the diffuse $H_{\rm I}$), and 3) cold dust (as in molecular
clouds). In the models of Devriendt et al.~\cite{dev1}, these
components are modelled as a mixture of polycyclic aromatic
hydrocarbon molecules (PAH), very small grains, and big grains. Big
grains may be either cold ($\sim$ 17 K), or heated by radiation from
star-forming regions (as suggested by observations of typical local
starburst galaxies like M82). A set of template spectra is then
constructed for galaxies of varying IR luminosity, with admixtures of
the various components selected in order to reproduce the observed
relations between IR/sub-mm color and IR luminosity. A similar
approach was used by \cite{dwek:98}, using a mixture of a typical
Orion-like $H_{\rm II}$ spectrum and an $H_{\rm I}$ spectrum
constructed to fit DIRBE observations of the diffuse ISM
\cite{dwek:97}. Here, we use the more empirical emission templates of
\cite{dwek:98} (kindly provided in electronic form by E. Dwek), but we
obtain very similar results with the models of \cite{dev1}.

The recipes for star formation, feedback, chemical evolution, and dust
optical depth contain free parameters, which we set for each model
(see SP) by requiring an average fiducial ``Milky Way'' galaxy to have
a K-band magnitude, cold gas mass, metallicity, and average B-band
extinction as dictated by observations of nearby galaxies.

Figure 1a shows the global star formation rate density for the two
star formation recipes that we consider here. The ``fiducial'' model
is the collisional starburst (CSB) model favored by SPF, in which
bursts of star formation may be triggered by galaxy collisions. The
``Late Star Formation'' model is the Constant Efficiency Quiescent
(CEQ) model of SPF, in which cold gas is converted to stars only in a
quiescent mode with constant efficiency. This produces a star
formation history similar to the models of the Durham
group\cite{baugh}, in which the
peak in the star formation history occurs at a more recent
epoch ($z \sim 1.5$) than in the CSB model. 
For the CSB model, we consider three different choices of IMF: Scalo
\cite{scaloIMF}, Salpeter \cite{salpeterIMF}, and Kennicutt
\cite{kenIMF}. These IMFs are graphed in Figure~\ref{fig:imf}.  For
the CEQ model we show only the Kennicutt case.  There is a noticable
difference in the far-UV and the mid- to far-IR.  The Scalo IMF
produces less UV light relative to optical and near-IR light, compared
to the Kennicutt and Salpeter IMFs, which produce more high mass
stars than the Scalo IMF, and thus more ultraviolet light to be
absorbed and re-radiated by dust in the far IR. In Fig.~1b we show the
redshift evolution of the far-UV ($2000 \AA$) luminosity density for
these four models, compared with observations.\footnote{These models
were also compared with the observed luminosity density from nearby
galaxies, obtained by integrating the luminosity functions of galaxies
resolved in recent redshift surveys at wavelengths ranging from 0.2 to
2.2 $\mu$m, in Fig. 3 of~\protect\cite{PGam2000}.  However, the SAM
outputs graphed there were inadvertently multiplied by a factor of
$h^4 \approx 0.25$.  In \cite{veritas}, we renormalized all the models
by requiring that they all agreed with the K-band point at 2.2 $\mu$m.
Here we do not do this since our current SAMs \protect\cite{sp,spf}
use a corrected version \protect\cite{shethtormen} of the
Press-Schechter formalism.}  The Scalo model falls short at all
redshifts, and the CEQ model, which agrees at $z=0$, falls short at
higher redshifts.\footnote{The Durham type models~\protect\cite{baugh}
of SPF also predict that LBGs have higher stellar mass than
observations indicate~\protect\cite{papovich}, while predicted stellar
masses from the CSB model of SPF are in good agreement with the
observations\protect\cite{venice}.} It is encouraging that our very
simple model for dust extinction, which we normalized in the B-band at
$z=0$, appears to yield the appropriate level of dust extinction in
the UV at higher redshifts (SPF).

\begin{figure}[t!] 
\centerline{\psfig{file=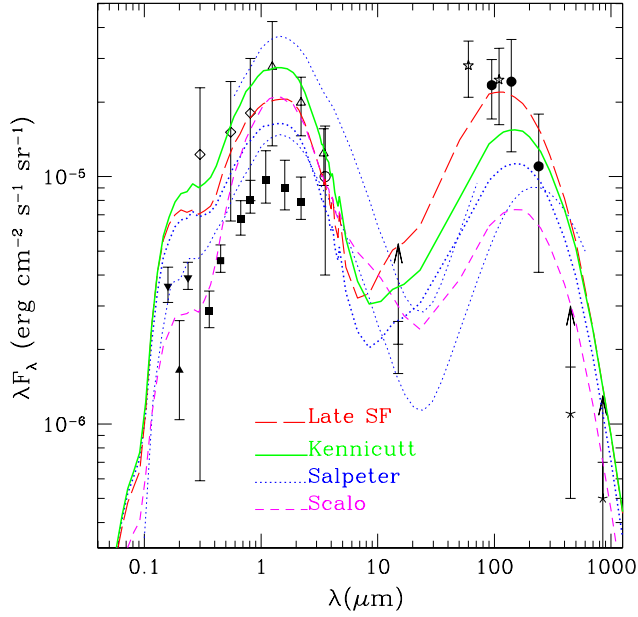,width=3.4in}}
\vspace{10pt}
\caption{Extragalactic background light: models and data. The far-UV
points are from STIS (inverted filled triangles)
\protect\cite{gardner} and FOCA observations (filled triangle)
\protect\cite{foca}.  The lower optical points (filled squares) are
lower limits from resolved sources \protect\cite{madaupoz}; the upper
ones (open diamonds) are from absolute photometry
\protect\cite{bernstein}.  The near-IR points are from DIRBE: (open
circle) \protect\cite{dwekarendt}, (open triangles)
\protect\cite{wright}.  The point at 15 $\mu$m is from ISOCAM resolved
sources \protect\cite{elbaz99}, and is thus a lower limit.  The far-IR
points are from DIRBE (filled circles) \protect\cite{hauser,lagache},
(stars) \protect\cite{fds}.  The curves are our results from modelling
the history of star formation in the \lcdm\ cosmology using
semi-analytic methods: a model with both quiescent star formation with
constant efficiency and starbursts, with Kennicutt, Salpeter, and
Scalo IMFs, and a Late SF model with only quiescent star formation with
constant efficiency (CEQ).  The lower light dotted curve is the
\lcdm\ EBL calculated using our previous methods
\protect\cite{veritas} for the Salpeter IMF, and the upper one is the
same curve to 80 $\mu$m multiplied by 2.5 for comparison with Mrk 501
data as analyzed by \protect\cite{guy} (see text).  Note that
$10^{-6}$ erg s$^{-1}$ cm$^{-2}$ sr$^{-1}$ = 1 \nwm2sr. }
\label{fig:ebl}
\end{figure}

\begin{figure}[t!] 
\centerline{\psfig{file=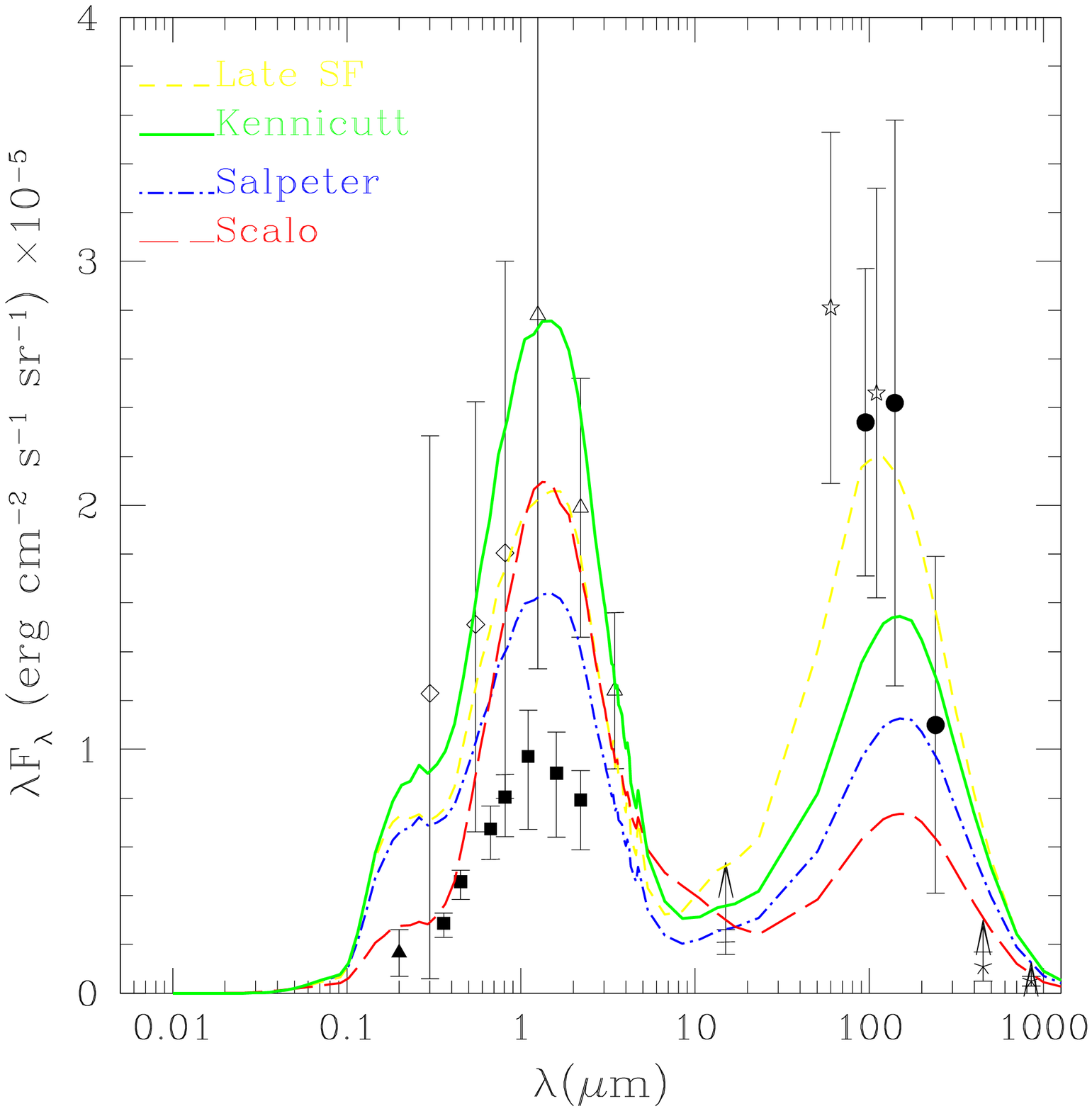,width=3.5in}}
\vspace{10pt}
\caption{Extragalactic background light: models and data with linear
vertical axis.  Labels are as in Fig. 3.}
\label{fig:ebl_linear}
\end{figure}

Recently, improved luminosity functions (LFs) have become available in
optical bands from the Sloan Digital Sky Survey\cite{blanton} and in
the K-band from 2MASS \cite{2masslumfcn,2dF_Klumfcn}.  We have found
that our CSB model with Kennicutt IMF agrees well with all of these
LFs when the average baryon fraction is $f_b=0.1.$ This
model is also consistent with the number counts in the mid-IR (15
$\mu$m from ISOCAM~\cite{elbaz99}) and far IR (60 $\mu$m from IRAS,
175 $\mu$m from ISOPHOT) but not the sub-mm (850 $\mu$m from
SCUBA~\cite{blain}). The resulting EBL is similar to that from the
Salpeter model discussed below.

\section{The Integrated Extragalactic Background Light}

Figure~\ref{fig:ebl} shows the EBL produced by our four models,
obtained by integrating the light over redshift (out to $z=4$) with
the appropriate K-corrections due to cosmological redshifting. We
compare this with a compilation of observational limits and
measurements of the EBL.  Fig.~\ref{fig:ebl_linear} presents the same
four models and the same data, but with a linear rather than
logarithmic vertical axis so that one can integrate the total energy
in the EBL by eye.  It is apparent that there is at least as much
energy in the far-IR part of the EBL as in the entire optical and
near-IR bands.  For example, Puget and collaborators \cite{puget}
estimated that the total energy in the EBL is between 60 and 93
\nwm2sr, with between 20 and 41 \nwm2sr\ contributed by the optical
and near-IR, and between 40 and 52 \nwm2sr coming from the far-IR.  If
the possible detection of the EBL at 60 $\mu$m by Finkbeiner et
al.~\cite{fds} were correct, that would further increase the far-IR
EBL; however, it is very difficult to determine the EBL at 60 $\mu$m
since the zodiacal light is so much brighter at that wavelength, and
it was probably partly confused with the EBL~\cite{finkIAU}.

The total energy in the EBL in units of critical density $\rho_c$ is
$\Omega_{\rm EBL} = (4\pi/c) (I_{\rm EBL}/\rho_c c^2) =
2.5\times10^{-8} I_{\rm EBL} h^{-2}$, where $I_{\rm EBL}$ is in units
of \nwm2sr.  The total energy density in the EBL corresponding to the
lower and upper estimates of \cite{puget} is $\Omega_{\rm EBL} =
(3.6-5.5) \times 10^{-6} (h/0.65)^{-2}$.  Although the EBL includes
energy radiated by active galactic nuclei (AGNs) as well as stars, it
is unlikely that AGNs contributed more than a few percent of the
total.  This is because the total energy radiated by AGNs is $E_{\rm
EBL}^{\rm AGN} = \eta \rho_{\rm BH} c^2$, where the efficiency of
conversion of mass to radiated energy in AGNs is $\eta \sim 0.05$.
Correspondingly, $\Omega_{\rm EBL}^{\rm AGN} = \eta \Omega_{\rm BH}
(1+z_{\rm BH})^{-1} \approx 4.5 \times 10^{-8} h^{-1} (\eta/0.05)
[3/(1+z_{\rm BH})] \lsim 0.02 \Omega_{\rm EBL}$.~\footnote{Updating
\cite{madau}, we have estimated $\Omega_{\rm BH}= (M_{\rm BH}/M_{\rm
spheroid}) \Omega_{\rm spheroid} \approx (1.5\times10^{-3})(1.8
\times10^{-3} h^{-1})$, using the observed (loose) correlation
\cite{kor} between a black hole mass and that of the galactic spheroid
in which it is found, and the estimated cosmological density of
spheroids \cite{fuku}.  Note that the factor $(1+z)^{-1}$ arises
because of the dilution of the contribution of high-redshift sources
due to the expansion of the universe.}  So for simplicity, in this
paper we will neglect the contribution of AGNs to the EBL.

Several interesting features emerge from the comparison of our SAM
models with the EBL data. In the UV to near-IR, the models are closer
to the direct measures of the EBL obtained by
\cite{bernstein,dwekarendt,wright} than to the lower limits from the
Hubble Deep Field~\cite{madaupoz}, although the Scalo IMF produces
less light in the UV because it has fewer high-mass stars.  As noted,
our Kennicutt CSB SAM with $f_b=0.1$, which agrees well with the
latest observed local luminosity density at $z=0$, produces an EBL
close to the Salpeter one in Fig.~4.  
Of our four new EBL curves, the Late SF model and the fiducial
Kennicutt model are also consistent with the DIRBE measurements at 140
$\mu$m.  The Salpeter EBL lies a little more than 2$\sigma$ below the
DIRBE measurement at 140 $\mu$m. The LateSF far IR is higher than the
other models because its later star formation suffers less dilution
due to the expansion of the universe.  The models differ significantly
in the mid-IR, $\sim10-60\mu$m, where the EBL can be probed by TeV
$\gamma$-rays.  The lower dotted curve in Fig.~\ref{fig:ebl},
representing our previous attempt \cite{veritas} to model the EBL, is
well below the 15 $\mu$m lower limit as well as the DIRBE measurements
at longer wavelengths.  As we stated in \cite{veritas}, we expected
our EBL results to change as we improved our dust emission modelling.
In addition to inclusion of the PAH features, the new dust emission
model has more warm dust than the one used in \cite{veritas}.

\begin{figure}[b!] 
\centerline{\psfig{file=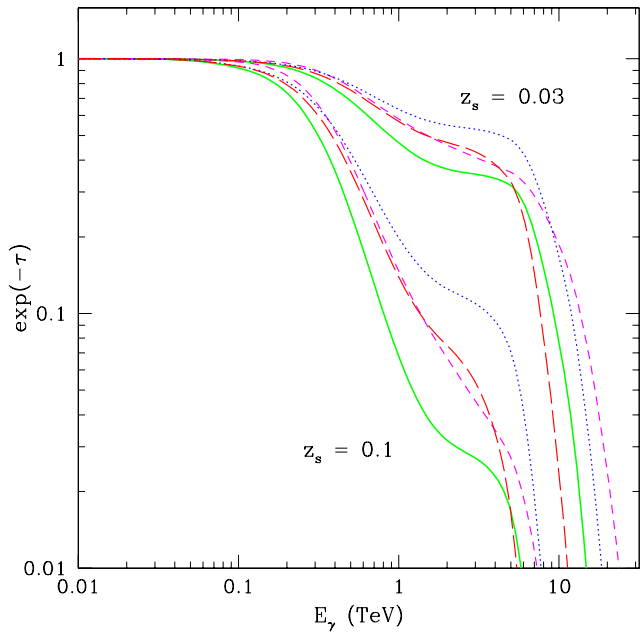,width=3.5in}}
\vspace{10pt}
\caption{The attenuation factor, $\exp(-\tau)$ for $\gamma$-rays as a
function of $\gamma$-ray energy for the four \lcdm\ models
considered in Fig. 4.  The assumed redshift of the source, z$_s$, is
indicated for each set of curves. 
}
\label{fig:atten}
\end{figure}

We now discuss constraints from the TeV $\gamma$-ray observations.

\section{Attenuation of high-energy $\gamma$-rays}

Figure~\ref{fig:atten} shows the $\gamma$-ray attenuation predicted by
the four \lcdm\ models considered here, for sources at redshifts
$z_s=0.03$ and 0.10.  All of the models predict rather little
absorption at $E_\gamma \lsim 5$ TeV for sources at $z_s=0.03$, but
fairly sharp cutoffs above $\sim5$ TeV, especially for the Late SF
model.  That model may be in conflict with the data from Mrk 501.  For
the blazars Mrk 421 and 501, both at $z\approx 0.03$, the synchrotron
self-Compton (SSC) model, in which $\sim$keV synchrotron X-radiation
from a very energetic electron beam is Compton up-scattered by the
same electrons to produce the observed $\sim$TeV $\gamma$-rays,
appears to explain both the keV-TeV spectra and their time variation
(see, e.g., \cite{guy,kraw99} and references therein).  Using a
simplified SSC model and keV X-ray data to predict the unattenuated
TeV spectrum of Mrk 501, Guy et al.~\cite{guy} used CAT and HEGRA data
to estimate the amount of $\gamma$-ray attenuation.  They find that
there is a rather good fit to the observed attenuation for the
\lcdm-Salpeter EBL from our earlier work~\cite{veritas} when it is
scaled upward by a factor of up to about 2.5 across the wavelength
range 1-80 $\mu$m; this is the upper Salpeter curve on
Fig.~\ref{fig:ebl}.  Our new Salpeter curve appears to be rather
consistent with this rescaling of our old Salpeter one, the Kennicutt
curve may be a little high, and the Late SF curve appears to be
definitely too high.  As noted earlier, the new Salpeter EBL curve in
Figs. 3,4 is similar to our latest $f_b=0.1$ Kennicutt SAM, which is
in good agreement with the latest local luminosity functions in the
optical and K bands from SDSS and 2MASS, and also in good agreement
with IR number counts from IRAS and ISO satellites.

The large flares in Mrk 501 in spring 1999 allowed an accurate
measurement of the gamma ray spectrum up to about 17 TeV, and
indicated that the spectrum had an exponential cutoff at about 5
TeV~\cite{Mrk501cutoff}. The flaring activity in Mrk 421 in early 2001
has now provided evidence for an exponential cutoff at about 4
TeV~\cite{krennrich}.  The coincidence in the cutoffs (within
observational uncertainties) for these two different extragalactic
sources suggests that both are due to absorption via pair production
on the EBL.  In order to confirm this, it will of course be necessary
to see similar cutoffs at lower energies for blazars at greater
distances.  There are already indications of this from 1426+42, a
blazar at redshift $z=0.13$ (four times farther than Mrk 421), on
which there is enough data from CAT~\cite{1ES1426},
Whipple~\cite{1H1426}, and HEGRA~\cite{felix} to begin to determine
the spectrum from a few hundred GeV to several TeV.  It will be very
useful to measure the spectrum from this and other sources at
comparable distances, such as PKS2155-304 at $z=0.116$, which will
soon be possible with the next generation of atmospheric Cherenkov
telescopes such as the H.E.S.S. array in Namibia, the CANGAROO-III
array in Australia, and the VERITAS array in Arizona.

Assuming that the EBL is like the LateSF curve in Figs. 3,4, several
authors (e.g.~\cite{promeyer}) have argued that the TeV attenuation
that this implies plus the observed spectrum of Mrk 501 leads to the
requirement that the spectrum at the source have a strong upturn in
its emitted flux above about 15 TeV, which would be very hard
(although perhaps not impossible~\cite{aharonetal}) to understand, and
they even suggest that one might have to abandon Lorentz invariance.
However, if one instead assumes our Salpeter EBL in Fig. 4 and
applies an SSC analysis to the HEGRA data from the Mrk 501 flares in
1997, the implied source spectrum is very reasonable, without an
upturn at the high-energy end~\cite{kraw01}.  

\begin{figure}[t!] 
\centerline{\psfig{file=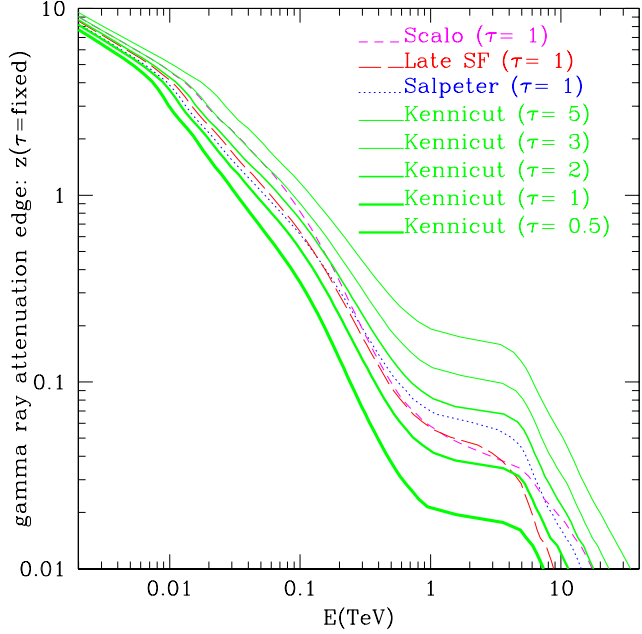,width=3.5in}}
\vspace{10pt}
\caption{The $\gamma$-ray attenuation edge.  The redshift where the
optical depth reaches unity is shown as a function of $\gamma$-ray
energy for each of the four \lcdm\ models considered in Fig. 4.  Also
shown for the Kennicutt IMF is the redshift where the optical depth
equals 0.5, 2, 3, and 5.}
\label{fig:edge}
\end{figure}

The compatibility of our new EBL calculations with the available data
on TeV $\gamma$-ray attenuation is definitely worth further
investigation.  The results appear to be sensitive to the details of
the models, raising the hope that they may be able to help answer
important questions about star formation and dust reradiation, and
also help to test the SSC modelling.  

Figure~\ref{fig:edge} depicts the $\gamma$-ray ``absorption edge,''
the redshift of a source corresponding to an optical depth of unity,
as a function of $\gamma$-ray energy.  Travelling through the evolving
extragalactic radiation field, $\gamma$-rays from sources at lower
redshift suffer little attenution.  The universe becomes increasingly
transparent as $E_\gamma$ decreases, probing the background light at
increasingly short wavelengths.  (We are using the treatment of
\cite{madau95} to account for absorption of ionizing radiation by the
Lyman alpha forest.)  The models all have the same qualitative
features, but differ significantly quantitatively.  The location of
the absorption edge is affected both by the assumed IMF and by the
history of star formation.  There is more absorption at most redshifts
with the Kennicutt IMF because with a higher fraction of high mass
stars, it is more efficient at producing radiation for a given stellar
mass; there is more absorption nearby in the Late SF model because the
starlight in this model is less diluted by the expansion of the
universe.  It is possible that measuring the transparency of the
universe to $\gamma$-rays at $\sim 100$ GeV with a number of sources
at various redshifts can provide a strong probe of star formation,
although there are uncertainties due to extinction by dust.

\section{Outlook}

The semi-analytic modelling of the EBL described here follows the
evolution of galaxy formation in time.  Forward modelling is a more
physical approach than backward modelling (luminosity evolution).
Pure luminosity evolution (e.g., \cite{ms98,ms00,steckeriau}) assumes
that the entire evolution of the luminosity of the universe arises
from galaxies in the local universe just becoming brighter at higher
redshift by some power of $(1+z)$ out to some maximum redshift.  It
effectively assumes that galaxies form at some high redshift and
subsequently just evolve in luminosity in a simple way.  This is at
variance with hierarchical structure formation of the sort predicted
by CDM-type models, which appears to be in better agreement with many
sorts of observations.

An alternative approach to modelling the EBL has been followed by Pei
and collaborators \cite{peifall95,fallcp,pei99}, in which they find an
overall fit to the global history of star formation subject to
constraints from input data including the evolution of the amount of
neutral hydrogen in damped Ly$\alpha$ systems (DLAS).  Their first
attempt \cite{peifall95,fallcp}, which was used as the basis for EBL
estimates by \cite{dwek:98,salamonstecker}, was somewhat misled by the
sharp drop in the DLAS hydrogen abundance from redshift $z\sim3$ to
$z\sim2$ reported in \cite{lwt95}.  With more complete data on DLAS
(see, e.g., Fig. 14 of \cite{slw00}) the $z=3$ point is lower and the
neutral hydrogen abundance is almost constant from $z=2$ to 4.  The
latest paper by Pei et al.~\cite{pei99} takes a variety of recent data
into account.  Their approach is to follow the evolution of the total
mass in stars, interstellar gas, and metals in a representative volume
of the universe; they assume a Salpeter IMF.  By contrast, the
semi-analytic methods we use follow the evolution of many individual
galaxies in the hierarchically merging halos of specific CDM models,
here \lcdm.  Despite the differences in approach, and the fact that
\cite{pei99} assumed $\Omega_m=1$ and Hubble parameter $h=0.5$, their
results are broadly similar to those from the semi-analytic approach
(see their \S4.4).  In particular, their EBL is similar to our old
results \cite{veritas} for the Salpeter IMF.  Our EBL results
presented here are higher in the near-IR and more consistent with the
direct determinations \cite{dwekarendt,wright}; they are also higher
in the mid-IR, probably mainly because of the warm dust and PAH
features in our dust emission model.  It will be interesting to see
whether further development of the global approach of Pei et al. and
of the semi-analytic approach lead to convergent results.

As our calculations show, the EBL, especially at $\lsim 1$ $\mu$m and
$\gsim 10$ $\mu$m, is significantly affected by the IMF and the
absorption of starlight and its reradiation by dust, as well as by the
underlying cosmology.  The cosmological parameters are becoming
increasingly well determined by other observations.  As data become
available on $\gamma$-ray emission and absorption from sources at
various redshifts, especially from the new generation of Atmospheric
Cherenkov Telescopes and the new $\gamma$-ray satellites AGILE and
GLAST, these data and their theoretical interpretation will help to
answer fundamental questions concerning how and in what environments
all the stars in the universe formed.

\section*{Acknowledgments}

I thank my collaborators Rachel S. Somerville, James S.  Bullock, and
Julien E. G. Devriendt.  My work was supported by NASA and NSF grants
at UCSC.  I am grateful for a Humboldt Award, and I thank Leo
Stodolsky for hospitality and Eckart Lorenz for enlightening
discussions about $\gamma$-ray astronomy at the Max-Planck-Institut
f\"ur Physik, M\"unchen.  I also thank Simon White for hospitality at
MPI Astrophysics in Garching, and Heinz V\"olk for hospitality at MPI
Nuclear Physics in Heidelberg.  I thank Henric Krawczynski, Paolo
Coppi, and Felix Aharonian for very helpful discussions of their SSC
modelling.  I thank Aldo Morselli for inviting me to give these
lectures at the ISSS, and for his patience in waiting for me to send
him the written versions.

\end{document}